\def\PEZZAG{9912025}
\documentclass{article}		
\usepackage{latexsym}		  

\begin{document}
\begin{titlepage}

\title{ \Large\bf The Mathematics of the Spinning Particle Problem
\thanks{ Based on presentation Available at:
Http://www.clifford.org/\~{}terler/talks/ } }
\author{ {\Large\bf Theodore G. Erler IV}
\thanks{Email:terler@physics.ucsb.edu} \\
\\ Department of Physics
\\ University of California
\\ Santa Barbara, CA }
\maketitle
\thispagestyle{empty}

\begin{abstract}
We introduce an original approach to geometric calculus in which we
define derivatives and integrals on functions which depend on extended
bodies in space--that is, paths, surfaces, and volumes etc. Though
this theory remains to be fully completed, we present it at its
current stage of development, and discuss it's connection to physical
research, in particular its application to spinning particles
in curved space.

{\it Summary of talk given at 
{\bf 5th International Conference on Clifford Algebras
and their Applications in Mathematical Physics},
Ixtapa-Zihuatanejo, MEXICO, June 27-July 4, 1999.}

\end{abstract}

\end{titlepage}


\def\SWITCH#1#2{#1}       

\def\HBAR{{\mathchar'26\mkern-9muh}}  
\def\vector#1{\vec{\bf {#1}}}         
\def\norm#1{\parallel{#1}\parallel}   
\def\LLL{{\cal L}}                    
\def\EQN#1{eq.\ (#1)}                 
\def\EB#1{{\bf e}_{#1}}
\def\EBH#1{{\bf \hat{e}}_{#1}}
\def\HALF{\mathstrut{1 \over 2}}                
\def\ODOT#1{\stackrel{\circ}{{#1}}}   
\def\ODOTS#1#2{\ODOT{#1}{\!}^{#2}}
\def\ODOTL#1#2{\ODOT{#1}{\!}_{#2}}
\def\VARO#1#2{\delta\!\!\ODOT{#1}{\!}^{#2}}

\def\COV#1{\nabla\!_{#1}}             
\def\PV#1#2{{\delta {#1} \over \delta {#2} }}
\def\PD#1#2{{\partial {#1} \over \partial {#2}}}
\def\PP#1#2#3{{\partial^2 {#1} \over \partial {#2} \partial {#3}}}
\def\BD#1#2{{\left[\partial_{#1},\partial_{#2}\right]}}
\def\CD#1#2{\left[ \partial_{#1}, \partial_{#2} \right]}
\def\LLL{{\cal L}}

\def\RRR{{\bf \underline{R}}}
\def\PPP{\underline{P}}
\def\DDD{{\bf d}_{\bf v}}                
\def\DAREA{{\bf d}^2_{\bf v \wedge w}}   
\def\LIM#1#2{\left .{d \ \over dx}\right |_{#1}\!\!\!\!\!\!\!\!  {#2} }
\def\FFF{F(\RRR+ x {\bf v})}
\def\VVV{\bf v}
\def\BACKNINE{\!\!\!\!\!\!\!\!\!}
\def\BACKSIX{\!\!\!\!\!\!}
\def\DVOL{{\bf d}^3_{\bf u \wedge v \wedge w}}   
\def\DFOUR{{\bf d}^4_{\bf t\wedge u \wedge v \wedge w}}
\def\XXX#1{{\bf x}_{#1}}

\pagestyle{myheadings}

\section*{I. Introduction}

\subsection*{A. Statement of Purpose}
In this paper, we wish to present a generalization of calculus that
has been developed in connection with the investigations into the
motion of spinning particles in curved space by
Pezzaglia\cite{Pezz97, Pezz98, Pezz9902, Pezz9903}.
The purpose of this calculus is twofold: First,
to provide some mathematical and conceptual support for Pezzaglia's
interpretation of his theoretical results. Secondly, to recast the notions
of differentiation and integration in a manifestly geometric way,
therefore suggesting new possible avenues for pure and applied
mathematical research. 

The basic idea of this new calculus is quite simple, though its
implications are broad: The idea is, simply,
to {\it define calculus} on functions
which depend upon extended objects in space--that is, paths, surfaces,
volumes etc. Usually, such functions are considered outside the domain
of possible consideration in calculus. For example, in Electrostatics,
we usually define the scalar potential as the mechanical work per charge:
$$V(\RRR) = \int_P {\bf E}({\RRR}^\prime)\cdot d{\bf v}^\prime \ .
\eqno(1.1)$$
Though this is a path integral, the potential is still only position
dependent because $\nabla \wedge {\bf E}=0\ $in electrostatics. However,
in electro{\it dynamics} $\nabla \wedge {\bf E}$ is not zero, and the
work integral \EQN{1.1} becomes path dependent. Since we do not
understand the calculus of path-dependent functions, we are forced to
abandon \EQN{1.1} in favor of the ``vector potential'' concept--even
though the work integral seems more physically relevant and intuitively
clearer (see, for example, Griffiths\cite{Griffiths}). Perhaps,
with a more complete calculus of paths and surfaces etc., we won't
need to be so theoretically constrained.

In this paper, we would like to trace the evolution of this new
calculus from its origins in Clifford Algebra, physics, and
multivariable calculus, to its current state of development. At this
time, we understand path dependent calculus relatively well, and are
still working to find a satisfactory generalization to surfaces and
volumes etc. Nevertheless, we have sufficient results to justify this
calculus as a possibly interesting avenue for further research, and
we hope to encourage more ideas in this direction.

Unfortunately, because of the limited space, we revert to stating our
results without proof. A more in-depth version of this paper is
available for the interested reader\cite{Erler}.

\subsection*{B. Clifford Algebra and Dimensional Democracy}
Many would argue that Clifford Algebra is perhaps the most authentic
algebraic representation geometry. Like no other theory, Clifford
Algebra integrates all of the dimensions of space into a seamless
whole, treating each dimension as equally important. Yet, at the same
time, Clifford Algebra has another distinct and powerful asset: it is
an extraordinarily useful language for expressing and understanding
the laws of physics, from gravitation to quantum mechanics. One might
wonder if this is not a coincidence, but, perhaps, the result of a
deep congruence between the content of physical law and the geometric
structure of Clifford Algebra. This line of reasoning has lead
Pezzaglia\cite{Pezz9902} to propose a new physical principle, which he
calls {\bf Dimensional Democracy:} {\it The structure behind the laws
of physics is reflected in the geometric language of Clifford Algebra.
Thus, physical law must itself completely and coherently utilize all
of the dimensions of space, as does Clifford Algebra.}

Or, more concisely,  {\it All of the dimensions within space are
equally important in determining the laws of physics.\/}
It is important to understand that ``Dimensional Democracy'' is a
{\it physical principle}, and as such, may be proven incorrect.
``Dimensional Democracy'' not only asserts that that Clifford Algebra
is {\it useful} in physics, but furthermore asserts that Clifford
Algebra's geometric philosophy is {\it necessary} for a full
understanding of physical law. 

\subsection*{C. The Spinning Particle Problem}
``Dimensional Democracy'' is a very broad principle, and can obviously
be taken in many directions. Using ideas inspired by Dimensional
Democracy, a successful Lagrangian derivation of the Papapetrou
equations for the motion of spinning particles in curved space, a long
standing problem in classical physics, has been recently
solved\cite{Pezz9903}. In this derivation, Pezzaglia\cite{Pezz98} makes
use of a particular corollary to ``Dimensional Democracy,''
which can be stated as follows:
{\it In the most general class of physics problems, each
dimension within space contributes through its own physical coordinate.}

To understand the motion of spinning particles in curved space, it is
necessary to use {\it both} a position and area coordinate, where the
position is conjugate to the particle's momentum and the area is
conjugate to its spin. This idea alone leads almost directly to the
Papapetrou equations\cite{Papapetrou, Pezz9902, Pezz9903}.
To see this, consider a particle in the absence of forces,
$$ 0 = {d \over d\tau} p^\mu \EBH{\mu}=\dot{p}^\mu \EBH{\mu} +
p^\mu {d \EBH{\mu} \over d\tau}\ . \eqno(1.2)$$ 
If the particle had no spin, we would derive the geodesic equation as
usual by invoking the chain rule,
$${d \EBH{\mu} \over d\tau} = {d x^\nu \over d\tau}
{d \EBH{\mu} \over dx^\nu}
= \dot{x}^\nu \Gamma_{\mu\nu}^{\ \sigma} \EBH{\sigma}\ . \eqno(1.3)$$
However, this is not correct for a spinning particle, because now we
have area coordinates as well as position coordinates; therefore we
need a more complete chain rule, one which includes derivatives with
respect to the area as well as position:
$${d \EBH{\mu} \over d\tau} =  {d x^\nu \over d\tau} \PD{\EBH{\mu}}{x^\nu}
+\HALF {d a^{\sigma\delta} \over d\tau}  \PD{\EBH{\mu}}{a^{\sigma\delta}}=
\dot{x}^\nu \Gamma_{\mu\nu}^{\ \sigma} \EBH{\sigma}
+\HALF \dot{a}^{\sigma\delta}
\PD{\EBH{\mu}}{a^{\sigma\delta}}\ .\eqno(1.4)$$
This is interesting, but what does it mean to take a derivative with
respect to area? As it turns out, to derive the Papapetrou equations, it
is necessary to propose that the area derivative is actually a
commutator of two position derivatives\cite{Pezz97}:
$$\PD{\ \ }{a^{\mu\nu}} \equiv \PD{\ }{x^\mu} \PD{\ }{x^\nu}
-\PD{\ }{x^\nu} \PD{\ }{x^\mu} = \left[ \PD{\ }{x^\mu} , 
\PD{\ }{x^\nu} \right] \ . \eqno(1.5)$$
With this definition, the area derivative of the basis vector becomes
a curvature tensor, and we find:
$$0= \left( \dot{p}^\mu  + p^\sigma \dot{x}^\delta
\Gamma_{\sigma\delta}^{\ \mu}  + \HALF \dot{x}^\gamma
S^{\sigma\delta} R_{\sigma\delta\gamma}^{\ \ \ \mu}
 \right) \EBH{\mu} \ .\eqno(1.6)$$
This is one of the Papapetrou equations\cite{Papapetrou}. One's reaction
may be mixed. This derivation only supports the notion of ``Dimensional
Democracy'' insofar as it is, somehow, geometrically {\it appropriate}
to define the area derivative as a commutator derivative. A primary
motivation for this paper is to affirm that, indeed, this definition
is very appropriate, provided one views calculus in a properly
geometric way.

\section*{II. Review of Position-Dependent Calculus}
For the purpose of developing a more suggestive and appropriate notation,
we begin by re-deriving and re-expressing familiar results from
multivariable calculus. Thus, the next few sections will not be new, but
will establish basic themes which will continue to guide us as we
consider more complicated types of functions later.

Multivariable calculus is concerned with functions which map a {\it point}
in an n-dimensional Euclidean space to an element from a linear vector
space, $V$.  The exact nature of $V$ will not be important for us here,
except that addition and multiplication by scalars must be defined on it.
Also, because we would like to construct {\it vectors} within our
Euclidean space, we imagine that each axis in the space is supplied with a
basis vector. Thus, we can represent the location of any point by using a
position vector $\RRR$. (The underline signifies that the vector is
to be interpreted as a displacement with respect to the origin --that
is, {\it not} a free vector.) So, in standard notation, we are
considering all functions $F(\RRR)$ such that
$F:E^n \rightarrow V;\RRR \mapsto F(\RRR)$,
where $E^n$ is the set of all Clifford vectors in the n-D Euclidean
space (as in Baylis\cite{Baylis}).

\subsection*{A. The Displacement Derivative}
We now seek some way to quantify the {\it change} in $F$ as a function
of $\RRR$. To this end, we define the {\it displacement derivative},
$\DDD : F \rightarrow V; F(\RRR) \rightarrow
\DDD F(\RRR)$,
$$d_{\bf v} F(\RRR) \equiv \epsilon \lim_{\alpha \to 0}
\alpha^{-1} \Bigl( {F(\RRR + \alpha {\bf v}) - F(\RRR)}
\Bigr)\ .\eqno(2.1)$$
In the above equation, ``$\epsilon$'' is an extremely small
quantity-an infinitesimal.  Just stating this will be sufficient for
our purposes,
but the interested reader will note that our treatment of infinitesimals
is essentially in keeping with the notions of Non-Standard
Analysis\cite{Robinson}.

The reason why we care about infinitesimals like ``$\epsilon$'' is that
they allow a very simple interpretation of the displacement derivative,
$\DDD$.  To see this, we note, 
$$F(\RRR + \epsilon {\bf v}) - F(\RRR) = d_{\bf v} F(\RRR) +
O(F\epsilon^2 |{\bf v}|) \ . \eqno(2.2)$$
Thus, $\DDD$ computes the leading infinitesimal change in
$F(\RRR)$ that results from a slight change of $\RRR$ in the
direction of ${\bf v}$.  
	The displacement derivative $\DDD$ has one other
particularly important feature: it is {\it linear} in its vector
argument. This property is called ``internal linearity'', which we
express by the equation,
$$ {\bf d}_{\bf v+w} = \DDD + {\bf d}_{\bf w} \ , \eqno(2.3a)$$
$$ {\bf d}_{c{\bf v}} = c\,\DDD \ ,\ c \in \Re \ .\eqno(2.3b)$$
Internal linearity reveals the most crucial insight (and assumption)
of calculus; that locally, functions appear {\it linear}.

\subsection*{B. Integrals and the Fundamental Theorem}
We now define the path integral, a higher-dimensional analogy to the
usual $\int f(x) dx$ in single variable calculus. Mainly what's new in
n-dimensions is that we are no longer constrained to integrate along a
straight line, but are free to move about in space while we integrate.
So, now, we integrate along a curved path described by the parametric
equation,
$$\RRR = \RRR(t) \ , \ \hbox{For \ }t
\in \left[ 0,T \right] \ . \eqno(2.4)$$
Given such a path, we can construct a ``path integral operator'',
which we symbolize as $\int_{\RRR(t);t\in[0,T]}$.  This operator is
defined on a function, $K(\RRR,{\bf v})$, which depends both on
position, $\RRR$ and the rate of change in position, ${\bf v}$.
Thus, $ K: E^n \times E^n \rightarrow V;(\RRR, {\bf v}) \mapsto
K(\RRR, {\bf v})$.
We define the path integral of $K$ as a mapping,
$$\int_{\RRR(t); t \in [ 0 , T ]}: V \rightarrow V;K(\RRR,{\bf v})
\mapsto \int_{\RRR(t); t \in [ 0 , T ]}
K(\RRR^\prime,{\bf v}^\prime) \ ,  \eqno(2.5) $$
which can be computed by the Reimann sum:
$$\int_{\RRR(t); t \in [ 0 , T ]}K(\RRR^\prime, {\bf v}^\prime)\equiv
{1 \over \epsilon} \lim_{NT \to \infty}N^{-1}
{ \sum_{n=0}^{T/\alpha}
K\left(\RRR(n\alpha), \left. {d \RRR \over dt}\right|_{t=n\alpha} \right)
}  \ . \eqno(2.6)$$ 
Now that we've defined both the displacement derivative and the path
integral by equations (2.1) and (2.6), we combine the definitions to
form the fundamental theorem of calculus:
$$\int_{\RRR(t);\ t \in [0,T]} \DDD F(\RRR^\prime) =
F[\RRR(T)] - F[\RRR(0)] \ . \eqno(2.7)$$
Of course, there
are other fundamental theorems in n-dimensions--the so-called Stoke's
theorems, which integrate over surfaces, volumes etc.
Hestenes\cite{Hestenes} showed that we could write all such
fundamental theorems in the form:
$$\int_{X^k} \left(d{\bf x}^{\prime k} \cdot \nabla \right)
F(\RRR^\prime)
=\oint_{\delta X^k} d{\bf x}^{k-1} F(\RRR^\prime)\ ,\eqno(2.8)$$
where $\int_{X^k}$ indicates integration over some k-dimensional body
and $\oint_{\delta X^k}$ indicates integration around the surface of
that body. For the most part, we will be concerned only with the path
integral fundamental theorem, except to note one thing: all fundamental
theorems as in \EQN{2.8} integrate in steps of one. That is, it relates
an integral over a k-surface to an integral over a k-1 surface, and the
integral of a single derivative $\nabla$ to an integral with no
derivatives. As we consider path and surface dependent functions, we will
find fundamental theorems which, amazingly, do not integrate in
steps of one! 

\subsection*{C.  Hints of a Path Dependent Calculus}
In single variable calculus, the fundamental theorem usually comes in
two distinct forms, one with the derivative {\it inside} the integral
and one with it {\it outside}:
$$\int^x_{x_0} {df \over dx^\prime} dx^\prime \equiv
f(x) -f(x_0) \ ,
\quad\quad {d\ \over dx} \int^x_{x_0} f(x^\prime) dx^\prime
\equiv f(x) \ . \eqno(2.9ab)$$
The first equation (2.9a) is basically the path integral theorem,
\EQN{2.7}, expressed in one dimension. The second one (2.9b), on the
other hand, doesn't appear to have any analogue in higher dimensions,
at least not that we've discussed. We could extrapolate this equation,
however; perhaps the generalization could look something like,
$$\DDD \int_{\RRR(t); t \in [0,T] } \!\!\!\!\! F(\RRR^\prime)
=F[\RRR(T)] \ . \eqno(2.10)$$
However, a moment's thought reveals that this apparently natural equation
has no meaning! The displacement derivative is defined on functions of
position, {\it not} on path dependent functions like
$\int F$, so \EQN{2.10} is effectively meaningless. For this reason,
mathematicians have usually presumed that there is no generalization of
the fundamental theorem with the derivative {\it outside} the integral.
However, this is not the only line we could take on this issue, perhaps
we should, instead, {\it define} differentiation on path dependent
objects so that theorems like \EQN{2.10} make sense. 

Further indication that we might need a path dependent calculus comes
from the Pezzaglia's ``area'' derivative, \EQN{1.5}. It is simple to
show that displacement derivatives always commute when they act on
position dependent functions:
$$\left[ \DDD, {\bf d}_{\bf w} \right] F(\RRR) = 0 \ .  \eqno(2.11)$$
Unfortunately, this is bad news for the area derivative, because
\EQN{1.5} defines it to be precisely the commutator derivative, which
is always zero! Clearly something more is needed, and there is reason
to think that a path-dependent calculus might solve this problem. In
particular, consider the following: The displacement derivative
$\DDD$ describes the change in $F$ through a displacement
which is bounded by two points. If the action of $\DDD$ is to
be nonzero, clearly the function must depend on position, otherwise
the function would not be able to detect the change between endpoints.
By analogy, perhaps the area derivative describes the change in
$F$ through an {\it area which is bounded by two paths.\/} Therefore,
$F$ must be at least path dependent for a nontrivial area derivative. 

\section*{III. Derivatives in Path Dependent Calculus}
We devote the next few sections to extending the definition of the
derivative to path dependent functions. The concepts and notation are
slightly more involved here, but the added complication is compensated
for by increased mathematical richness, which has yet to be fully
explored. In part IV we will define integration on path dependent
functions as well, from which we will be able to derive new and
interesting fundamental theorems.

\subsection*{A.  Representing a Path}
This section is devoted to developing a useful notation for
representing paths as mathematical objects. We could, of course,
describe paths in the normal way by thinking of them as parameterized
sets of points as described in \EQN{2.4}.
Unfortunately, this representation is too difficult to work with for
our purposes. So, let us propose a more practical notation based on
an operation called {\it stacking}.  Stacking, symbolized by $\Rightarrow$,
is basically the process of bringing two objects together to form an
ordered pair. Thus,
$a \Rightarrow b = (a,b), \ \ \hbox{for}\  a,b \in S$,
where ``$S$'' is some arbitrary set. Really, stacking is no more than a
new symbolism for designating ordered pairs, but it portrays ordered
pairs in a more dynamic fashion, as an {\it operation} which is
performed between two objects, rather than as a completed object in
itself.
Any object which is created by stacking other objects together is
called a {\it stack}. Notice that stacking is {\it not} communitive;
this property will be what allows us to define non-commuting
derivatives, as we will soon see.

Of particular interest to us is are stacks composed of
{\it vectors} in $E^n$, for example,
$$s = {\bf u}_1 \Rightarrow {\bf u}_2 \Rightarrow 
\dots \Rightarrow {\bf u}_n \ . \eqno(3.1)$$
For the purpose of visualization, we imagine that stacks such
as \EQN{3.1} look like sequences of vectors which have been ``glued'' on
top of one another, head to tail, like a structure of rigid rods in
space. The motivation behind this visualization is that we can create
a continuous path by simply stacking together an infinite sequence of
infinitesimal vectors.  Thus, if we are given a parametric equation
$\RRR=\RRR(t), t\in[0,T]$, we can write a path,
$$ P \equiv \lim_{NT \to \infty}
\sum^{NT}_{n=1}\!\!\Rightarrow
\left[ \RRR(n/N) - \RRR([n-1] / N) \right] \ .  \eqno(3.2)$$
In the above equation, the $\sum \Rightarrow$ symbol simply denotes
a ``summation'' carried out with arrows rather than pluses, (which,
incidentally, means we have to be careful to carry out the ``sum'' in
the specified order). As \EQN{3.2} is written, the path $P$ doesn't have
any particular location within space, but we can define its location by
adjoining it with a position
vector, $\underline{P} = \RRR(0) \Rightarrow P$, which fixes the
starting end of the path to the position $\RRR(0)$.
Given a path \PPP, we would like to define an operation that
allows us to break it into smaller pieces. Thus, we define the
``partition'' of a path, symbolized ${}_a(P)_b$,
$$ {}_a(P)_b \equiv \lim_{NT \to \infty}
\sum^{Nb}_{n=Na}\!\!\Rightarrow
\left[ \RRR(n/N) - \RRR([n-1] / N) \right] \ .  \eqno(3.3)$$
We assume that $0\leq a \leq b \leq T$.  In the special cases
$a=0$ or $b=T$, we will simply write
${}_0(P)_b=(P)_b$ and ${}_a(P)_T={}_a(P)$.

To shorten our equations later, we symbolize a straight line path
in the direction of a vector {\bf v} by,
$$L_T {\bf v} = \lim_{NT \to \infty}\!\!\!\sum_{n=1}^{NT}
\!\!\Rightarrow (n/N){\bf v} \ . \eqno(3.4)$$

\subsection*{B.  The Displacement Derivative}
We now extend the definition of a the displacement derivative so that
it can act on {\it path dependent} functions. By ``path dependent
function'', we mean any mapping, $G$, which takes a path
$\underline{P}$ into an element of a linear vector space $V$.
Thus, denoting the set of all paths in $E^n$ by $\Phi^n$, we can
define $G$ as a function,
$G:\Phi^n \rightarrow V; \ \underline{P} \mapsto G(\underline{P})$.
Notice that we have underlined the path $\underline{P}$ to remind us
that its location in space has been fixed.

Now, when we take the derivative of a path dependent function, we do not
vary the path in an arbitrary way, rather we choose a particular point
along the path we want to vary, and then change the path at this point
{\it only}. This may seem restrictive, but if we want to compute a
more general variation, we can ``add up'' derivatives taken at different
points along the path. So, before we can define the derivative, we have
to invent a notation that will tell us which point along the path
$\underline{P}$ we are allowed to change as we differentiate. We
signify this by placing an asterisk `$\star$' at the chosen point
along the path, like so:
$\underline{P} = \underline{P}_i^\star \Rightarrow {P}_f$.  Here
$\underline{P}_i$ is the part of the path preceding the point
of differentiation (in the stacking sense), and
${P}_f$ is the part after. 

We are now ready to consider the displacement derivative, which we
define as the mapping,
${\bf d}_{\bf v}: V \rightarrow V; \ G(\underline{P})
\mapsto {\bf d}_{\bf v} G(\underline{P})$
such that,
$${\bf d}_{\bf v} G(\underline{P}_i^\star \Rightarrow
P_f ) \equiv \epsilon \lim_{\alpha \to 0}
\alpha^{-1} \bigl[G(\underline{P}_i^\star \Rightarrow
L_\alpha {\bf v} \Rightarrow P_f ) 
-G(\underline{P}_i^\star \Rightarrow P_f )
\bigr] \ , \eqno(3.5)$$
where $L_\alpha {\bf v}$ is a straight line path in the direction
of ${\bf v}$, as in \EQN{3.4}.  By making a Taylor expansion, we
can see that this definition has a simple interpretation:
$$G\left(\underline{P}_i^\star \Rightarrow L_\epsilon {\bf v}
\Rightarrow {P}_f \right) - G(\underline{P}_i^\star
\Rightarrow {P}_f ) = {\bf d}_{\bf v}
G(\underline{P}_i^\star \Rightarrow {P}_f ) +O(\epsilon^2 G|{\bf v}|)
\ . \eqno(3.6)$$
Thus, ${\bf d}_{\bf v}$ computes the first order change in $G$ that
results when we insert an infinitesimal path $L_\epsilon {\bf v}$ 
between the paths $\underline{P}_i$ and $P_f$.

Before we can accept \EQN{3.5} as an acceptable derivative, we must
demonstrate that ${\bf d}_{\bf v}$ satisfies external linearity and
obeys a Leibniz product rule. The external linearity of
${\bf d}_{\bf v}$ is trivial, and a product rule is easily derived:
$$\DDD c( \underline{P}) G(\underline{P})
=c( \underline{P})\  \DDD G(\underline{P}) +
G(\underline{P})\ \DDD c( \underline{P}) \ , \eqno(3.7)$$
for $c \in \Re; \ G \in V$. 
Therefore, \EQN{3.5} defines an acceptable derivative operator.

The question remains as to whether the displacement derivative
satisfies internal linearity, \EQN{2.3}, when it acts on path
dependent functions (that is, whether path dependent
functions are geometric). 
Unlike the position dependent case, we cannot provide any
reasonably general proof of internal linearity for path dependent
functions. Therefore, we restrict ourselves from the outset to
considering only path dependent functions which are geometric. 

This is great, but are there any geometric path dependent
functions? Indeed, there are. For example, the line integral,
$$G( \PPP ) = \int_{\PPP } A( \RRR^\prime ) d{\bf v} \ , \eqno(3.8)$$
allows internal linearity to be satisfied (assuming that
$A(\RRR)$ is geometric). Is this the only example? Thankfully, no,
because we can use \EQN{3.8} as the ``seed'' from which we can
create an infinite variety of new geometric functions. In particular,
we can take any number of line integrals, combine them using any
smoothly varying function we like, and the result will still be
geometric. Clearly, we could create a huge number of different
functions this way, so we can see that internal linearity is not
too stringent a condition to impose.

\subsection*{C.  The Area Derivative}
As we saw earlier, there is strong reason to think that path dependent
calculus may possess the key to understanding the area derivative, and
also to providing some support to Pezzaglia's geometric
derivation of the Papapetrou equations\cite{Pezz9902, Pezz9903}.
In this section, we see how far we can go
towards realizing these goals. Thus, we define the area
derivative $\DAREA$ as a mapping,
$\DAREA:V\rightarrow V;G(\PPP)\mapsto \DAREA G(\PPP)$,
given by the limit,
$$\DAREA\,G(\PPP_i^\star \Rightarrow P_f) \equiv$$
$$\epsilon^2 \lim_{\alpha \to 0} \alpha^{-2} \Bigl[
G\left(\PPP_i^\star \Rightarrow L_\alpha {\bf v} \Rightarrow
L_\alpha {\bf w} \Rightarrow P_f \right) - 
G\left(\PPP_i^\star \Rightarrow L_\alpha {\bf w} \Rightarrow
L_\alpha {\bf v} \Rightarrow P_f \right) \Bigr]. \eqno(3.9)$$
This definition may strike the reader as dubious, because both
vectors ${\bf v}$ and ${\bf w}$ are individually needed to compute
the limit, whereas $\DAREA$ should depend only on the
area ${\bf v\wedge w}$, by hypothesis. We will discuss this problem
in a moment, but for now let's see how to interpret \EQN{3.9}.  By
making a Taylor expansion, we can derive:
$$G(\PPP_i^\star \Rightarrow L_\epsilon {\bf v} \Rightarrow
L_\epsilon {\bf w} \Rightarrow P_f)-
G(\PPP_i^\star \Rightarrow L_\epsilon {\bf w} \Rightarrow
L_\epsilon {\bf v} \Rightarrow P_f) = $$
$$\DAREA G(\PPP) + O(\epsilon^2 G|{\bf v}| 
|{\bf w}| ). \eqno(3.10)$$
This tells us that the area derivative computes the lowest order
change in $G$ that occurs between two paths that enclose a little
parallelogram of area $\epsilon^2 {\bf v \wedge w}$.  Thus, the area
derivative encapsulates the idea of change through a small area, just
as the displacement derivative describes change through an
infinitesimal displacement.

If the proposition \EQN{1.5} is correct, it would seem that we should
be able to write our area derivative \EQN{3.9} as the commutator of
two displacement derivatives. In fact, it is simple to show that we can:
$$\DAREA = \DDD {\bf d}_{\bf w} - {\bf d}_{\bf w} \DDD
= \left[ \DDD, {\bf d}_{\bf w} \right] \ . \eqno(3.11)$$
Thus, we've succeeded at explaining why the derivative with respect
to area should be a commutator derivative. For the purpose touching
base with standard theory, we re-express \EQN{3.11} using the
familiar notation of tensors and covariant derivatives:
$$\left[ \nabla_\mu , \nabla_\nu \right] V^\alpha =
R_{\mu\nu\beta}^{\ \ \ \ \alpha}\, V^\beta \ . \eqno(3.12)$$
This is a well-known result from differential geometry (without
torsion), and can be derived straightforwardly from \EQN{3.11}. We
note, however, that \EQN{3.11} is actually more general than
\EQN{3.12}, because it applies to every geometric path dependent
function. Equation (3.12), on the other hand, only applies to vector
functions in curved space.

We now consider the pressing question of whether or not the area
derivative, as given by the limit \EQN{3.9}, is really a well defined
object. To see why we should be concerned, consider the two area
derivatives: $\DAREA G(\PPP) $ and
${\bf d}^2_{[{\bf v}+a{\bf w}]\wedge [{\bf w} + b({\bf v} + a
{\bf w})]} G(\PPP)$. 
It is clear that these two derivatives should be the same, because
they differentiate with respect to the same area:
$$[{\bf v}+a{\bf w}]\wedge [{\bf w} + b({\bf v}+a{\bf w})]=
[{\bf v}+a{\bf w}]\wedge {\bf w} + b [{\bf v}+a{\bf w}]
\wedge [{\bf v}+a{\bf w}] $$
$$ ={\bf v\wedge w} + a {\bf w \wedge w}
={\bf v \wedge w} \ . \eqno(3.13)$$
However, if we take definition \EQN{3.9} literally, we should
compute $\DAREA$ and ${\bf d}^2_{[{\bf v}+a{\bf w}]\wedge [{\bf w}
+ b({\bf v} + a{\bf w})]}$ using two completely different limits. Unless
by some miracle these limits happen to converge to the same value, we must
abandon the area derivative concept to avoid contradiction. 
In fact, for a general path dependent function, these two limits will be
different, and we cannot define the area derivative. However, if the
function is {\it geometric\/}, they will not be different.
The easiest way to demonstrate this is to rewrite the limits
as commutator derivatives, invoke internal linearity, and sort out
the terms. Therefore, the area derivative is well defined as long
as it acts on a geometric function.	

There are two things we should check before we accept \EQN{3.9} as
an acceptable derivative: that it acts linearly on functions and obeys
the Leibniz product rule.  The area derivative is clearly linear,
$$\DAREA [G(\PPP) + H(\PPP)] =
\DAREA G(\PPP) + \DAREA H(\PPP) , \eqno(3.14a)$$
$$\DAREA\,c\,G(\PPP) = c\,\DAREA G(\PPP) , \eqno(3.14b)$$
for $c \in \Re$ and $G, H \in V$.  The product rule follows directly from
the area-commutator derivative relation \EQN{3.11},
$$\DAREA c(\PPP)\,G(\PPP) = c(\PPP)\,\DAREA
G(\PPP) + G(\PPP)\,\DAREA c(\PPP)\ , \eqno(3.15)$$
for $c \in \Re$ and $G \in V$.
Therefore, the area derivative is an acceptable differential operator.

We hope that the reader has been convinced that \EQN{3.9} describes
what should rightly be called the ``area derivative'', not only because
it is consistent with the area derivative of \EQN{1.5}, but because it
seems to be the uniquely natural extension of the derivative concept
from a 1-D ``displacement'' to a 2-D ``area''.

\subsection*{D.  The Volume Derivative}
The reader is doubtless familiar with the fact that the derivative of
a constant is zero,  
$$\DDD c = 0 \ . \eqno(3.16)$$
We have also seen that the area derivative of a position dependent
function is always zero,
$$\DAREA F(\RRR) = 0 \ . \eqno(3.17)$$
So, by analogy, we might think that there is a ``volume'' derivative,
${\bf d}^3_{\bf u \wedge v \wedge w}$ which gives zero when it acts on a
path dependent function:
$${\bf d}^3_{\bf u \wedge v \wedge w} G(\PPP) = 0 \ . \eqno(3.18)$$
Equations (3.16-18) show an interesting geometric progression, but we
need to define the volume derivative before this trend can be concretely
realized. 

So, let's imagine how we might quantify the change in a path dependent
function with respect to a volume. This attempt may seem suspect, because
changing a path will sweep out a surface, not a volume. However, perhaps
we can have a path sweep out a closed surface which {\it bounds} a
volume, and define the resulting change to be the volume derivative.
This means, specifically, adding up the changes which occur as we move
the path across each face of the parallelepiped of
volume ${\bf u \wedge v \wedge w}$:
$$\matrix{{\bf d}^3_{\bf u \wedge v \wedge w} G(\PPP)\equiv 
&\ \;(\hbox{change across right side})
&+(\hbox{change across top})\quad\quad\ \cr
&+(\hbox{change across front side})
&+(\hbox{change across left side})\ \ \cr
&+(\hbox{change across bottom})\ \ 
& + (\hbox{change across back side})\cr
} $$
or more precisely,
\def\AWFUL#1#2#3{G\left(\left({\bf #1},{\bf #2}\right),{\bf #3}\right)}
\def\AWFUR#1#2#3{G\left({\bf #1},\left({\bf #2},{\bf #3}\right)\right)}
\def\OAWFUL#1#2#3{G[\underline{P}_i\Rightarrow\!(L_\alpha {\bf #1}
\Rightarrow\!{L_\alpha{\bf #2}})\Rightarrow\!{L_\alpha{\bf #3}}
\Rightarrow\!{P_f}]}
\def\OAWFUR#1#2#3{G[\underline{P}_i\Rightarrow\!{L_\alpha{\bf #1}}
\Rightarrow\!{(L_\alpha{\bf #2}}\Rightarrow\!{L_\alpha{\bf #3}})
\Rightarrow\!{P_f}]}
$${\bf d}^3_{\bf u \wedge v \wedge w} G(\PPP)\equiv$$
$$\epsilon^3 \lim_{\alpha \to 0} \alpha^{-3} \left( \matrix{
\phantom{+}\AWFUL{w}{u}{v} - \AWFUL{u}{w}{v} 
+\AWFUR{w}{v}{u} - \AWFUR{w}{u}{v} \cr
+\AWFUL{v}{w}{u} - \AWFUL{w}{v}{u} 
+\AWFUR{v}{u}{w} - \AWFUR{v}{w}{u} \cr
+\AWFUL{u}{v}{w} - \AWFUL{v}{u}{w} 
+\AWFUR{u}{w}{v} - \AWFUR{u}{v}{w} \cr
}\right) \eqno(3.19a)$$
$$\hbox{where}: \AWFUL{x}{y}{z}\equiv \OAWFUL{x}{y}{z} \ , \eqno(3.19b)$$
$$\hbox{and\ \ }:\AWFUR{x}{y}{z} \equiv \OAWFUR{x}{y}{z} \ . \eqno(3.19c)$$
At a glance, we realize that all these terms add to precisely zero. Thus
it seems that, though we can move a path around a volume in our
imagination, path dependent functions won't know about it! To get a
nontrivial volume derivative, we have to step beyond path dependent
functions  and consider {\it surface} dependence. 

It is straightforward to show that each term in \EQN{3.19} can be
reproduced by computing,
$${\bf d}^3_{\bf u \wedge v \wedge w} =\left[{\bf d}^2_{\bf u \wedge w},
{\bf d}_{\bf w} \right] + \left[{\bf d}^2_{\bf v \wedge w},
{\bf d}_{\bf u} \right] + \left[{\bf d}^2_{\bf w \wedge u},
{\bf d}_{\bf v} \right] \ . \eqno(3.20)$$
Surprisingly, this equation directly implies that {\it displacement
derivatives don't associate on surface dependent functions.\/} When we
speak of ``non-associativity'', we do not mean it in the usual sense
of ordering operations. In other words,
$({\bf d}_{\bf u}{\bf d}_{\b v}){\bf d}_{\bf w}$
does not mean ``first take
${\bf d}_{\bf u}{\bf d}_{\bf v}$ and then multiply
${\bf d}_{\bf w}$ from the left''; this doesn't really make sense in the
context of differentiation. What we mean is that, for any string of
three derivatives,
${\bf d}_{\bf u}{\bf d}_{\bf v}{\bf d}_{\bf w}$, two of them must
be ``taken together'', in some appropriate sense, in order
for ${\bf d}_{\bf u}{\bf d}_{\bf v}{\bf d}_{\bf w}$ to be meaningful.

\section*{IV.  Integrals in Path Dependent Calculus}
	With the definition of the displacement derivative, \EQN{3.5}, we
have come enormously far: We have explained non-commutivity of
derivatives, come to understand the area and volume derivatives,
made contact with differential geometry and the Papapetrou equations,
and caught a glimpse of the curious non-associativity of surface
dependent calculus. However, what eventually justifies the displacement
derivative is not these things, but our ability to use it to create new
and useful fundamental theorems. Thus, we devote the next few sections
to defining integration and fundamental theorems for path dependent functions.

\subsection*{A.  Path Integration}
We define the path integral as a mapping of a path dependent function,
$$\int^P_{\underline{P_0}} \!\!\!:V \rightarrow V;G(\PPP)
\mapsto \int^P_{\underline{P_0}} \!\!\!G(\PPP^\prime)\ , \eqno(4.1)$$
given by the Riemann sum,
$$\int^P_{\underline{P_0}} \!\!\!G(\PPP^\prime)\equiv
\left({1 \over \epsilon}\right)
\lim_{NT(P)\to \infty}\!\!N^{-1} \sum^{NT(P)}_{n=0} 
G\left[\underline{P_0}\Rightarrow (P)_{n/N} \right] \ , \eqno(4.2)$$
where $(P)_{n/N}$ is a partition of $P$ as defined in \EQN{3.3}. 

With this definition, we can form two distinct fundamental theorems.
The first has a displacement derivative {\it inside} the path integral,
$$\int^P_{\underline{P_0}} \!\!\!\DDD G(\PPP^{\prime\star})=
G(\PPP_0 \Rightarrow P) - G(\PPP_0) \ , \eqno(4.3)$$
where ${\bf v}$ is the tangent vector at the end of the path $\PPP^\prime$.
This is the direct generalization of our earlier path integral
theorem, \EQN{2.7}. The second fundamental theorem, as we alluded to in
\EQN{2.10}, has a derivative {\it outside} the path integral:
$$\DDD\int^{P^\star}_{\underline{P_0}} \!\!\! G(\PPP^{\prime})=
G(\PPP_0 \Rightarrow P^\star) \ . \eqno(4.4)$$
This result is more than a coincidence; in fact, the definition of $\DDD$
\EQN{3.5} was largely chosen so that this theorem would work out. So,
in a sense, all of path dependent calculus has it's origin in \EQN{4.4}.
Furthermore, \EQN{4.4} encourages us to ponder the existence of higher
dimensional theorems,
$$\DAREA \int\BACKNINE\int\limits_{\smash{surface}}\BACKSIX H = H,
\quad \DVOL\int\BACKSIX \int\limits_{\smash{volume}}\BACKSIX\int I = I,
\quad etc.\quad \dots \ .\eqno(4.5)$$
Though we don't yet fully understand the calculus of surface and
volume dependence etc., we can use these equations to guide further
generalization; just as \EQN{4.4} guided the development of path
dependent calculus. 

\subsection*{B.  Surface Integrals}
The two fundamental theorems we have discussed so far, equations
(4.3) and (4.4), both involve the displacement derivative. What about the
area derivative? Does it have any  fundamental theorem? In fact, it does, a
fundamental theorem involving the {\it surface} integral of an area
derivative. Before we can express this theorem, however, we have to define
what it means to integrate a {\it path dependent} function over a surface.

We would like to think of a surface as a parameterized set of paths,
$$\PPP = \PPP(s);\quad s\in[0,S] \ , \eqno(4.6)$$
where the paths all share the same endpoints and parameterization
interval, $T$. Therefore, we can imagine a surface to be ``sheet''
swept out by a continuum of paths bound between two points,
$\RRR_i$ and $\RRR_f$.
Alternatively, we could represent a surface as a set of points
parameterized by two variables, $(s,t)$:
$$\RRR = \RRR(s,t);\quad s\in [0,S], \ t\in [0,T] \ , \eqno(4.7)$$
such that $\RRR(s,0)$ and $\RRR(s,T)$ are both fixed points in space,
independent of $s$.

Having said this, we define the surface integral on a function,
$K(\PPP,{\bf v, w})$ which depends both on a path $\PPP$ and the
{\it change in} $\PPP$ through two vectors, $({\bf v,w})$, which
can be visualized as tangent to the surface of change. Thus, the
surface integral operator is a mapping,
$$\int\BACKNINE\int\limits_{\PPP(s);s\in[0,S]}\BACKNINE:V \rightarrow
V;K(\PPP, {\bf v, w}) \mapsto
\int\BACKNINE\int\limits_{\PPP(s);s\in[0,S]}\BACKNINE K(\PPP,{\bf v,w})
\ , \eqno(4.8)$$
given by the double Riemann sum,
$$\int\BACKNINE\int\limits_{\PPP(s);s\in[0,S]}\BACKNINE\!\!\!K(\PPP,{\bf v,w})
\equiv \left({1 \over \epsilon^2}\right) \lim_{NT,NS \to \infty}\!\!\!N^{-2}
\sum\limits^{NS}_{m=1}\sum\limits^{NT}_{n=1}K\left[
\PPP(m,n),{\bf v}(m,n),{\bf w}(m,n)\right] \ , \eqno(4.9a)$$
\medskip
$$where:\quad \PPP(m,n)=\bigl(\PPP\left[m/N\right]\bigr)_{n/N}^\star
\Rightarrow {}_{n/N}\bigl(\PPP\left[m/N\right]\bigr)\ , \eqno(4.9b)$$
$${\bf v}(m,n)=\left.{d\RRR(m/N,t) \over dt}
\right|_{n/N} \ , \eqno(4.9c)$$
$${\bf w}(m,n)=
\left.{d\RRR(s,n/N) \over ds}\right|_{m/N} \ . \eqno(4.9d)$$
This equation looks formidable, but really the idea is quite
simple: we are adding up $K$ over all of the paths $\PPP(s)$
and over the changes which are occurring at each point along those
paths.  Combining the definition of the surface integral with the
area derivative, equations (3.9) and (4.9), we discover the
remarkable fundamental theorem,
$$\int\BACKNINE\int\limits_{\PPP(s);s\in[0,S]}\BACKNINE\DAREA
G(\PPP^\prime) = G[\PPP(S)]-G[\PPP(0)] \ . \eqno(4.10)$$
This result is immediately distinguished from other fundamental
theorems by one extraordinary feature: {\it it integrates in a
step of two\/}. This means, specifically, that it relates a
{\it double} integral over a {\it second order} derivative to
something with no integrals and no derivatives. By contrast,
fundamental theorems in standard theory always integrate 
{\it once} over a {\it single} derivative. This is true of
differential forms, whose fundamental theorem integrates over
{\it only one} exterior derivative\cite{Flanders},
$$\int\limits_\Sigma {\bf d} \omega =
\int\limits_{\partial\Sigma} \omega  \ . \eqno(4.11)$$
We would never see a fundamental theorem which integrates two
exterior derivatives, because such a theorem would be trivial by
the Poincare Lemma, ${\bf d}^2 \omega=0$. If we complete the
generalization to surface and volume dependence etc., we expect
to find even bigger fundamental theorems which integrate in $k$ steps:
$$\int\!\!\!\int\limits_{X^3}\!\!\!\int \DVOL H = \Delta H\,,
\quad
\int\!\!\!\int\limits_{X^4}\!\!\!\int\!\!\!\int \DFOUR H = \Delta H
\ , \eqno(4.12ab)$$
$$\int\!\!\!\int\dots\int\limits_{X^k}
{\bf d}^k_{\bf \XXX{1}\wedge\XXX{2}\wedge\dots\wedge\XXX{k}} H =
\Delta H \ . \eqno(4.12c)$$
Eq.(4.12abc) is, doubtless, an impressive geometric hierarchy of
fundamental theorems; but still, we need to do much more thinking
before \EQN{4.12} can be given any definite meaning. It is our
great hope that this profound generalization may some day be realized.

We should mention one other remarkable property of the area derivative
fundamental theorem: it automatically includes the 2-D Stoke's theorem
as a special case, as we can verify by evaluating \EQN{4.10} using a
line integral as in \EQN{3.8}. It is significant that we can derive the
2-D Stoke's theorem from the area derivative theorem, \EQN{4.10}.
In fact, perhaps {\it all} Stoke's theorems can be derived from suitably
generalized fundamental theorems, as in \EQN{4.12}. In this view,
Stoke's theorem, which was previously thought to be ``fundamental'',
is no more than a shadow of more powerful theorems from a more general
calculus.

\section*{V.  Conclusion}
\subsection*{A. Overlap with Standard Theories}
\subsubsection*{1. Differential geometry}
The author is aware that standard theory has already made contact
with path dependent calculus, at least partially, through the notion
of {\it curvature} in differential geometry. Indeed, this is why the
ideas in this paper have anything to do with the Papapetrou equations
for spinning particles in curved space. 

We should emphasize, however, that path dependent calculus actually
gives us more than differential geometry alone. In particular,
differential geometry is concerned with only one type of path
dependent function, a basis vector in curved space, while path
dependent calculus concerns {\it all} path dependent functions
(as long as it satisfies internal linearity). But also, differential
geometry makes it difficult to appreciate the significance of
{\it path dependence} as a powerful and logical extension of
multivariable calculus. This situation is aggravated by the definition
of the covariant derivative, $\nabla_\sigma$, which allows one to
calculate in curved space without any reference to path dependent
basis vectors. Thus, with the path dependent objects out of sight,
one sees no need for a calculus of path dependent functions.

\subsubsection*{2. The Calculus of Variations and Functional Analysis}
When people first hear about this calculus, they often think that it
must bear some close relationship to the calculus of variations and
functional analysis. Actually, the relationship is not as strong as
one might initially suppose, the two theories are capable of
describing each other, but not in a simple or natural way. 

Functional Analysis considers paths, surfaces, and volumes to be like
{\it points} in an infinite dimensional vector space. Hence, in this
view, a path dependent function is really no more than a function
of {\it position} in infinite dimensions. This concept allows us to
directly generalize the ideas of position dependent calculus to
functions of path, surface, and volume etc. Thus, a directional
derivative in infinite dimensions can be visualized as an arbitrary
path variation which could, if specifically chosen, describe the
displacement or area derivatives. However, this approach is not very
pleasing, the significance of the displacement derivative is drowned
out by a vast sea of other equally possible path variations which are
of no particular importance. We could, on the other hand, approach
this from the opposite direction, that is, describe a general path
variation as a sum of displacement and area derivatives. However,
it is not clear that we gain an appreciable amount for our efforts.
Thus, while path dependent calculus could be made equivalent to
Variational Analysis, and vice versa, it is not particularly
natural or illuminating to do so. 

\subsubsection*{3. The Geometric Calculus of Hestenes and Sobczyk}
There may be some question as to how much the calculus of this paper
has in common with the Geometric Calculus of Hestenes and Sobczyk,
after all, both theories define differentiation with respect to
vectors and bivectors etc., and both are, in some sense,
``geometric''. However, the two theories are essentially different
in their domain of consideration. Geometric Calculus concerns,
mostly, functions which depend on a general element of the Clifford
Algebra, that is, a multivector\cite{Hestenes}. Our calculus
concerns functions of paths and surfaces etc., and makes only
modest use of Clifford Algebra. Perhaps there is something to be
gained by trying to unify these two theories, but for now, it is
clear that they are not the same.

\subsection*{B. Where Do We Go From Here?}
\subsubsection*{1. Completing the Generalization}
In the author's opinion, the fate of this calculus depends crucially
on our ability to generalize it to functions of arbitrary
hyper-surfaces n-dimensions. Consistency of principle demands that,
if we did it for paths, we must be able to do it for surfaces and
volumes etc. However, the question is very complex, and the author
has not yet been able to find a satisfactory generalization. If a
general definition for the displacement derivative is ever found,
we expect it to satisfy the following properties:
\bigskip 

(i) It must be non-associative in a greater number of
parentheses when it differentiates functions of more
complex hyper-surfaces.

(ii) It must yield intuitive definitions for the area,
volume etc. derivatives through relationships such as
\EQN{3.11} and \EQN{3.20}.

iii) It must reduce to the position and path dependent
definitions of $\DDD$ as a special case.

iv) It must satisfy internal linearity for a
sufficiently broad class of functions.

v) It must allow us to write path, surface, volume etc.
integral fundamental theorems with the derivative on the
{\it outside}, as in \EQN{4.5}.

vi) It must also allow us to write path, surface, volume
etc. fundamental theorems with the derivative inside, as
in \EQN{4.12}.

\bigskip
\noindent
This is a lot of mathematical weight to be carried by a single
definition, so we must define $\DDD$ with great care. The author
is currently working on some ideas.

\subsubsection*{2. A New Approach to Differential Geometry}
It is quite striking how easily we were able to derive fundamental
results from differential geometry through a few modest insights
and simple definitions. Concepts such as curvature, torsion,
and the Bianchi identity which are usually deeply buried in
formalism, are cleanly derivable with path dependent calculus.
Perhaps differential geometry might benefit from a complete
pedagogical re-derivation within the framework of path dependent
calculus, a problem currently under investigation.

One of the interesting possibilities such a re-derivation might
offer is the prospect of generalizing the concept of space to
allow for {\it surface} dependent basis vectors. Thus, the Bianchi
identities would no longer vanish and the volume derivative would
give us a new fundamental tensor,
$${\partial^3 \EBH{\mu}\quad \over \partial V^{\alpha\beta\gamma}}
=U_{\alpha\beta\gamma\mu}^{\quad\quad\sigma}\,\EBH{\sigma}
\ . \eqno(5.1)$$
What this equation essentially means is that the concept of
parallel transport of a point along a path is replaced by the
concept of parallel transport of a {\it path through a surface\/}.
Possible applications to String Theory come to mind, but at this
point, the idea is far from being concretely realized.

\subsection*{C. Acknowledgement}
I would like to thank Dr. Bill Pezzaglia of Santa Clara University
for sharing his thoughts,
encouraging my ideas, working to get these ideas published
(n.b. help with typesetting), and for
his great and friendly concern for my development as a scientist.

\end{document}